\documentclass[twocolumn,preprintnumbers,amsmath,amssymb,pra]{revtex4}
\usepackage{epstopdf}
\usepackage[colorlinks,linkcolor=blue,anchorcolor=blue,citecolor=blue]{hyperref}
\usepackage{txfonts}
\usepackage{subfigure}
\usepackage{graphicx,hyperref}
\usepackage{dcolumn}
\usepackage{bm}
\usepackage{extarrows}
\usepackage{setspace}
\usepackage{float}
 \allowdisplaybreaks
 
\begin{document}


\title{Spin Josephson effects of spin-orbit-coupled  Bose-Einstein condensates in a non-Hermitian double well}
\author{Jia Tang$^{1}$}
\author{Zhou Hu$^{1}$}
\author{Zhao-Yun Zeng$^{2}$}

\author{Jinpeng Xiao$^{2}$}
\author{Lei Li$^{2}$}

\author{Yajiang Chen$^{1}$}
\author{Ai-Xi Chen$^{1}$}
\author{Xiaobing Luo$^{1,2}$}
\altaffiliation{Corresponding author: xiaobingluo2013@aliyun.com}
\affiliation{$^{1}$Department of Physics, Zhejiang Sci-Tech University, Hangzhou, 310018, China}
\affiliation{$^{2}$School of Mathematics and Physics, Jinggangshan University, Ji'an 343009, China}

\date{\today}

\begin{abstract}
In this paper, we investigate the spin and tunneling dynamics of a spin-orbit-coupled  noninteracting Bose-Einstein condensate in a periodically driven non-Hermitian  double-well potential. Under high-frequency driving, we obtain the effective time-averaged Hamiltonian by using the standard time-averaging method, and analytically calculate the Floquet quasienergies, revealing that the parity-time ($\mathcal{P} \mathcal{T}$)-breaking phase transition appears even for arbitrarily small non-Hermitian parameters when the spin-orbit coupling strength takes half-integer value, irrespective of the values of other parameters used. When the system is $\mathcal{PT}$-symmetric with balanced gain and loss, we find numerically and analytically that in the broken $\mathcal{P} \mathcal{T}$-symmetric regions, there will exist the net spin current together with a vanishing atomic current, if we drop the contribution of the exponential growth of the norm to the current behaviors.
When the system is non-$\mathcal{PT}$-symmetric, though the quasienergies are partial complex, a stable net spin current can
be generated by controlling the periodic driving field, which is accompanied by a spatial localization of the condensate in the well with gain. The results deepen the understanding of non-Hermitian physics and could be useful for engineering a variety of devices
for spintronics.\\

{Keywords:} spin-orbit coupling, current, non-Hermitian, $\mathcal{P} \mathcal{T}$-symmetric
\end{abstract}
\pacs{}
\keywords{:SO-coupling,current,non-Hermitian,$\mathcal{P} \mathcal{T}$-symmetric}
\textsl{}\maketitle
\section{introduction}
Over the years an intense research effort has been made to investigate non-Hermitian systems both theoretically and experimentally\cite{moiseyev2011non,el2018non}. In conventional quantum mechanics, the Hermiticity requirement of a Hamiltonian guarantees the reality of the energy spectrum and
conserved total probability. However, it is ubiquitous in nature that the
quantum probability (the norm of the state) is effectively not conserved due to the exchange of  energy,  particles,
and information between the environment and the
quantum subsystem  of our interest\cite{rotter2015review}. An important development in the physics of non-Hermitian systems
was the discovery, by Carl Bender and co-workers, that a broad
class of non-Hermitian Hamiltonians can exhibit purely real spectra as long as the system possesses a combined parity and time-reversal
symmetry, that is, parity-time ($\mathcal{PT}$) symmetry\cite{Bender1998, Bender2002}.  A distinctive characteristic  in $\mathcal{PT}$-symmetric systems is the spontaneous-symmetry-breaking phase transition, where the spectrum
changes from all real (exact phase) to  complex (broken phase) when the gain-loss coefficient exceeds
a critical threshold. The exploitation of $\mathcal{PT}$-symmetric systems with static (i.e., time-independent) potentials has
been prolific. Recently, manipulation of the spontaneous $\mathcal{PT}$-symmetry breaking (and non-Hermitian physics) by  making use of periodic driving schemes  has also attracted much
attention\cite{West2010, Moiseyev2011, Xiao2012, Valle2013, Luo2013, Luo2014, Luo20142, Lian2014, Zhong2016, Gong2015, Lee2015, Yang2016, Luo2017, Wu2017, hitsazi2017, Li2019, Xiao2017}.  For example, as
shown in Refs.~\cite{Luo2013, hitsazi2017}, the spontaneous-symmetry-breaking phase transition emerges even for arbitrarily weak gain-loss coefficient by adjusting the parameters of periodic driving field.

On another front, there have been remarkable  progresses and research activities in the study of the quantum dynamics of  Bose-Einstein condensates (BECs) in a double-well potential, due to the fundamental significance and numerous potential applications. The prototypical system of an atomic BEC in a double-well
potential represents  a bosonic Josephson junction (BJJ)\cite{Smerzi1997, Milburn1997}, i.e., a bosonic analogue of the well-known superconducting Josephson junction\cite{Josephson1964}, and the coherent dynamical behaviors such as the Josephson oscillations (JO) and macroscopic quantum self-trapping (MQST) have been observed experimentally\cite{Albiez2005, Levy2007}.
 Early theoretical efforts
have already been carried out in generalizing the Josephson junctions with scalar condensates to mixtures\cite{Ashhab2002, Sun2009, Satija2009, Ng2012},
or spinors\cite{wang2008spinor, mele2012spin, julia2009spinor, You2002, You2005, You2007}, and  a variety of fundamental tunneling phenomena have been uncovered. On the other hand, the controlled removal of atoms from a Bose-Einstein condensate
(BEC) was realized by using the experimental technique based on
the electron microscopy\cite{Gericke2006, Gericke2008}, which promotes the boson-Josephson junction manipulated by
a local dissipation
 as a governable open quantum system for implementing the switching between a self-trapping
state and the macroscopic quantum tunneling regime\cite{Shchesnovich2010}. In addition, experimental realization of a $\mathcal{PT}$-symmetric two-well system  of ultracold atoms  is made possible by
embedding it within additional time-dependent wells which act as particle reservoirs, as identified in the early proposal\cite{Kreibich2013, Kreibich2014}.

Spin-orbit coupling (SOC), the interaction
between the particle dynamics and its spin, has already been
extensively studied in diverse branches of physics, which contributes to  the electronic fine structure of atoms and condensed matter
phenomena and applications like  topological insulator\cite{hasan2010colloquium}, spin Hall effect\cite{Xiao2010}, and spintronic\cite{RevModPhys.76.323}.  In  cold atom systems,
spin-orbit coupling can be generated experimentally by coupling two hyperfine
states of atoms via a pair of counterpropagating Raman lasers\cite{lin2011spin}.   Such
Raman-induced spin-orbit coupling opens new possibilities for investigating the Josephson effects (JEs) in two-component cold atom  systems,  for which two components can be explained as two hyperfine
(pseudospin) atomic states.  Recently, the role played by SOC  on the tunneling dynamics of BECs in double-well
potentials were addressed in several works\cite{zhang2012josephson, 2014Josephson, PhysRevA.90.033618, citro2015spin, wang2015spin, kartashov2018dynamical, 2020Controlling, li2022physics}. In Ref.~\cite{zhang2012josephson}, Zhang and co-worker discovered that a net atomic spin current termed as spin JEs can be induced by the spin-dependent tunneling between two wells.
Subsequently, Ref.~\cite{2014Josephson}  concentrated on the
effect of atom-atom interactions and provided a classic study of  self-trapping dynamics of the spin polarization and
population imbalances of each  bosonic
pseudospin species. A parallel work analytically treated the quantum behavior of spin-orbit
coupled BECs from the viewpoint of a two-mode
Bose-Hubbard-like Hamiltonian\cite{citro2015spin}.
The dynamical suppression of tunneling of spin-orbit-coupled  noninteracting Bose-Einstein
condensate in a double-well potential under periodic driving were reported in Ref.~\cite{ kartashov2018dynamical}.
We also notice that two very similar  four-level systems were
studied in Refs.~\cite{khomitsky2012spin, li2018qubit}, which discussed the spin dynamics of a single spin-orbit-coupled electron in a double
quantum dot. In view of the aforementioned achievements both in non-Hermitian physics and spin-orbit-coupled BJJ, it is natural to ask the following two important questions: can the $\mathcal{PT}$-symmetry-breaking phase transition be induced for arbitrarily weak gain-loss coefficient if given certain suitably chosen values of SOC strength? can the net atomic spin current still occur in the non-Hermitian two-well system of cold atoms with SOC?

The aim of this paper is to answer the above concerns and questions by investigating the non-Hermitian system of a spin-orbit-coupled atom (or noninteracting Bose-Einstein
condensate) in a  periodically driven double-well potential.  Fortunately, the answers to both of the
questions above are definite ``yes". In such a system, we have the following main observations: (i)
managing effective SOC alone can achieve  $\mathcal{PT}$-symmetry-breaking transition for arbitrarily small non-Hermitian parameters, which has the same effect as the use of  periodic driving schemes; (ii)
despite existence of non-Hermiticity,  a net spin current (i.e., the atomic current is zero while the spin current is nonzero), which is termed as spin Josephson effects (JEs), can still exist, indicating that there are spin exchanges but no net-particle tunneling between the two wells of the potential.
\section{Model}
\begin{figure}[htbp]
\center
\includegraphics[width=9cm]{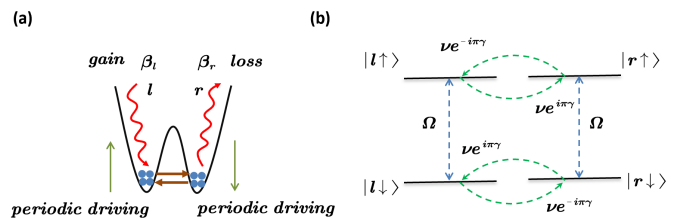}
\caption{(color online) Schematic view of (a) a spin-orbit-coupled noninteracting BEC in a double well.
Atoms are injected into the left well (gain) and removed from the right well (loss). The two wells of the potential are  out-of-phase
driven periodically.
(b) The dynamic process of
the system, where the solid lines represent the four modes and the dashed lines
represent the Raman coupling and the spin-dependent interwell tunneling.}
\label{figure1}
\end{figure}

In the pioneering experiment by NIST group\cite{lin2011spin}, the synthetic
SOC was successfully  realized by  coupling two hyperfine
states of atoms via a pair of Raman lasers. Relying on this experimental setup, we consider a single ultracold atom (or noninteracting BEC) with two hyperfine
pseudospin states $\left|\left.\uparrow\right>\right.$ and $\left|\left.\downarrow\right>\right.$ in a periodically driven open double-well potential with synthetic SOC. Assuming that the pseudospin bosonic atom occupies the lowest state of
each well, the quantum dynamics of a spin-orbit-coupled atom (or noninteracting BEC)  confined in a periodically driven open double well can be rather generally described by the non-Hermitian Hamiltonian\cite{2020Controlling, ji2019bloch}
\begin{align}\label{con:1}
	\hat{H}=&
	\sum_{\sigma}{\left[ \left( \varepsilon+i\beta _l \right)\hat{n}_{l\sigma}-\left( \varepsilon+i\beta _r \right) \hat{n}_{r\sigma} \right]}\nonumber\\&+\frac{\varOmega}{2}\sum_{j,\sigma}{\hat{c}_{j\sigma}^{\dag}\hat{c}_{j\sigma ^{'}}}-
	\nu\left( \hat{c}_{l}^{\dag}\hat{T}\hat{c}_r+H.c. \right).
\end{align}
 Here  $\hat{c}_{j}^{\dag}=\left( \hat{c}_{j\uparrow}^{\dag},~\hat{c}_{j\downarrow}^{\dag} \right)$, $\hat{c}_j=\left( \hat{c}_{j\uparrow},~\hat{c}_{j\downarrow} \right) ^{\textrm{T}}$ (superscript
$\textrm{T}$ stands for the transpose),  $\hat{c}_{j\sigma}^{\dag}$ ($\hat{c}_{j\sigma}$)
describes the
creation (annihilation) of a pseudospin $\sigma =\uparrow,\downarrow$ boson
in the $j$th ($j=l, r$) well, and  $H.c.$ denotes the
Hermitian conjugate of the preceding term. $\hat{n}_{j\sigma}=\hat{c}_{j\sigma}^{\dag}\hat{c}_{j\sigma}$
 represents the number operator for spin $\sigma$  in the $j$th  well, $\nu$ is the usual tunneling amplitude without SOC, and $\varOmega$ is the Raman coupling strength.
The  spin-dependent hopping matrix $\hat{T}=e^{-i\pi \gamma \hat{\sigma}_z}$ is obtained through Peierls substitution\cite{Peierls1933}, where $\hat{\sigma}_z$ is the $z$ component of Pauli operator, and $\gamma=2d/\lambda_r$ (here, $d$ is the distance of two trap centers and  $\lambda_r=2\pi/k_r$ is the
wavelength of the Raman laser) characterizes the effective SOC strength. The tunneling is controlled by out-of-phase periodic modulation of depths of two wells of the potential, described
by $\varepsilon=\alpha \cos \left( \omega t \right)$ with $\alpha$ the driving amplitude and $\omega$  the driving frequency. In addition
to the periodic driving, we incorporate gain and loss mechanisms into the system, and, without loss
of generality, assume the non-Hermitian coefficient $\beta_j>0$, which indicates that  the left well experiences gain while the right
well loss.  A schematic view of our
model system is shown in figure \ref{figure1}. Throughout this paper, we have set $\hbar=1$ and let the system parameters $\alpha,~\omega,~\beta_j,~\nu,~\varOmega$ be in units of the reference frequency $\omega _0=0.1E_r=2.25\text{kHz}$ with $E_r=\hbar^2k_{r}^{2}/\left( 2m \right)$ being the single-photon recoil energy, and time be normalized in units of $\omega_0^{-1}$. In the experiments\cite{Albiez2005, lin2011spin, Lignier2007, Spielman2015}, the system parameters
can be adjusted in a wide range: $\nu, \varOmega, \beta_j \sim \omega_0$ and $\alpha \sim \omega \in [0, 100](\omega_0)$.

 Since the Hamiltonian \eqref{con:1} is periodic in time with
period $\tau=2\pi/\omega$, the Floquet theorem
tells us that there exists a
complete set of solutions to the time-dependent Schr\"{o}dinger equation
$i\partial |\varPsi(t)\rangle/\partial t=\hat{H} |\varPsi(t)\rangle$  of the form
$\left| \left. \psi _n\left( t \right) \right> \right.=e^{-i\varepsilon _nt}|\varphi _n(t)\rangle,~|\varphi _n(t)\rangle =|\varphi _n(t+\tau)\rangle$, where $|\varphi _n \left( t \right) \rangle$ are Floquet states and $\varepsilon_n$ are quasienergies. The Floquet states can be obtained by solving the eigenvalue equation	
\begin{equation}\label{con:2}
\left( \hat{H}-i\frac{\partial}{\partial t} \right) |\varphi _n \left( t \right) \rangle =\varepsilon_n|\varphi _n \left( t \right) \rangle.
\end{equation}

In general, it is hard to obtain the exact Floquet solutions of Eq.~\eqref{con:2},
but the quantum dynamics can be
investigated analytically in high-frequency region where $\omega$ is far greater than all other
frequencies of the physical system. In the high-frequency limit, the effective time-averaged Hamiltonian can be derived by using a time-averaging method, which has been routinely employed in periodically driven quantum systems. According to the well-established method, a static effective Hamiltonian can be obtained by
time averaging of the periodic-driving terms, i.e.,
 \begin{align}\label{Heff}
\hat{H}_\text {eff} =&\frac{\omega}{2\pi}\int_0^{\frac{2\pi}{\omega}}{dt\hat{S}^{-1}\left[ \left. -\nu \left( \hat{c}_{l\uparrow}^{\dag}e^{-i\pi \gamma}\hat{c}_{r\uparrow}+\hat{c}_{l\downarrow}^{\dag}e^{i\pi \gamma}\hat{c}_{r\downarrow}+H.c. \right) \right] \right.}\hat{S}\nonumber\\&+
\frac{\omega}{2\pi}\int_0^{\frac{2\pi}{\omega}}{dt\hat{S}^{-1}}\left[ \sum_{\sigma}{\left( i\beta _l\hat{n}_{l\sigma}-i\beta _r\hat{n}_{r\sigma} \right) +\frac{\varOmega}{2}\sum_{j,\sigma}{\hat{c}_{j\sigma}^{\dag}}\hat{c}_{j\sigma ^{'}}} \right] \hat{S},
\end{align}
where $\hat{S}=e^{-iA \left( t \right) \sum_{\sigma}{\left( \hat{n}_{l\sigma}-\hat{n}_{r\sigma} \right)}}$ and $
A\left( t \right) =\int_0^t{\left[ \alpha \cos \left( \omega t \right) \right]}dt=\frac{\alpha}{\omega}\sin \left( \omega t \right)$. Implementing the integral in Eq.~\eqref{Heff}, we get the effective Hamiltonian
\begin{align}\label{effH2}
	\hat{H}_\text {eff} =&\sum_{\sigma}{\left( i\beta _l\hat{n}_{l\sigma}-i\beta _r\hat{n}_{r\sigma} \right)}\nonumber\\&+\frac{\varOmega}{2}\sum_{j,\sigma}{\hat{c}_{j\sigma}^{\dag}\hat{c}_{j\sigma ^{'}}}-\left( \hat{c}_{l\uparrow}^{\dag}J\hat{c}_{r\uparrow}+\hat{c}_{l\downarrow}^{\dag}J^{\ast}\hat{c}_{r\downarrow}+H.c. \right),
\end{align}
where $J=\nu e^{-i\pi \gamma}\mathcal{J} _0\left( \frac{2\alpha}{\omega} \right) $
with $\mathcal{J} _0\left( \frac{2\alpha}{\omega} \right) $ being the zeroth-order Bessel function of variable $\frac{2\alpha}{\omega}$.

We can solve the eigenvalue equation with the time-independent effective Hamiltonian \eqref{effH2},
\begin{equation}\label{eigenvalue}
\hat{H}_\text{eff}|\varphi _n'\rangle =E_n|\varphi _n'\rangle,
\end{equation}
where $|\varphi _n'\rangle$ and $E_n$ are eigenvectors and eigenvalues, respectively.

 Note that the unitary transformation operator $\hat{S}$  has the same period $\tau=2\pi/\omega$ as the Hamiltonian \eqref{con:1}. Thus, through the inverse transformation, we can construct the approximate expressions of Floquet solutions to  the original Hamiltonian \eqref{con:1} as follows
\begin{equation}\label{appF}
|\psi_n\left( t \right)\rangle=|\varphi _n\left( t \right) \rangle e^{-i\varepsilon _nt}=\hat{S}|\varphi _n'\rangle e^{-iE _nt},
\end{equation}
 where $|\varphi _n\left( t \right)\rangle=\hat{S}|\varphi _n'\rangle$ inherits the period of the driving force, satisfying $|\varphi _n\left( t+\tau \right)\rangle=|\varphi _n\left( t \right)\rangle$.
 This implies that $|\varphi _n\left( t \right)\rangle=\hat{S}|\varphi _n'\rangle$ are the so-called Floquet states and the
eigenvalues $E_n$ in Eq.~\eqref{eigenvalue} are the corresponding analytical quasienergies.

Taking the $\sigma$ Wannier state  $|j,\left. \sigma \right>=\hat{c}_{j\sigma}^{\dag}|0\rangle$ localized in the $j$th ($j=l, r$) well as basis, we expand the quantum state of system \eqref{con:1} as
$|\varPsi \left( t \right) \rangle =\sum_{j,\sigma}{c_{j\sigma}}\left( t \right) |j, \sigma \rangle$, where $c_{j\sigma}\left( t \right) $ indicates  the probability amplitude of finding a  pseudospin-$\sigma$ atom to be localized in the $j$th well. In the Wannier representation, the Hamiltonian operators can be represented in form of $4\times 4$ matrix.
By diagonalizing the effective Hamiltonian \eqref{effH2}, we get the eigenvalues
(approximate quasienergies) as
\begin{align}\label{con:9}
E_{1,~2}=\frac{1}{2}\left( i\beta _l-i\beta _r\mp \sqrt{m-w} \right),E_{3,~4}=\frac{1}{2}\left( i\beta _l-i\beta _r\mp \sqrt{m+w} \right),
\end{align}
where
\begin{align}\label{con:10}
&m=4\left[\nu\mathcal{J} _0\left( {2\alpha}/{\omega} \right)\right]^2+\varOmega^2-(\beta_l+\beta_r)^2,\nonumber\\&w=2\varOmega\sqrt{4\left[\nu\mathcal{J} _0\left( {2\alpha}/{\omega} \right)\right]^2\cos^2\left(\gamma \pi\right)-(\beta_l+\beta_r)^2}.
\end{align}

The analytical Floquet states and quasienergies provide basic
concepts and tools for treatment of the periodically driven system \eqref{con:1}, from which all available
time-dependent information about the system  can be deduced.  At any
time, the quantum state can be expanded  in the basis of  the Floquet eigenstates, namely,
\begin{align}\label{evqutumstate}
|\varPsi\left( t \right) \rangle =\sum_{n=1}^4{a_n}|\varphi _n\left( t \right) \rangle e^{-i\varepsilon _nt}
=\sum_{n=1}^4{a_n}\hat{S}|\varphi _n'\rangle e^{-iE_nt},
\end{align}
where $a_n$ are components of the quantum state, which are time-independent and determined by the initial state, i.e., $|\varPsi\left( t=0 \right) \rangle =\sum_{n=1}^4{a_n}|\varphi _n\left( t=0 \right) \rangle$.

\section{Currents in the $\mathcal{P} \mathcal{T}$-symmetric systems}
 First, we consider the situation of balanced gain and loss, where the loss (gain)
coefficients of two wells take the same values, $\beta _l=\beta _r=\beta$. In such situation, Hamiltonian
 \eqref{con:1} is $\mathcal{P} \mathcal{T}$ symmetric because of $\hat{P}\hat{T}\hat{H}=\hat{H}\hat{P}\hat{T}$,  where
the parity operator  $\hat{P}$ corresponds to the exchange of the two wells numbered by $l$ and $r$, and
 the time
reversal operator is defined as $\hat{T}: t+t_0\rightarrow -t+t_0$ ($t_0$ is an appropriate time point), $i\rightarrow -i$.

 In this work, our focus is placed on the current behaviors of  non-Hermitian system under action of SOC and periodic driving. To this end, we introduce the  population imbalance between the two wells
 \begin{equation}\label{pimbalance}
P_a=\left<  \varPsi \left( t \right) \right|  \varsigma\left|  \varPsi \left( t \right) \right>
=\langle \varPsi (t)|\hat{z}\otimes \hat{I} |\varPsi (t)\rangle,
 \end{equation}	
and  magnetization
 \begin{equation}\label{con:11}
P_s=\left< \left. \varPsi \left( t \right) \right| \right. \varLambda\left| \left. \varPsi \left( t \right) \right> \right.
=\langle \varPsi (t)|\hat{z}\otimes \hat{\sigma_z} |\varPsi (t)\rangle,
 \end{equation}	
where $\hat{z}$ denotes the position operator,  $\hat{z}=\sum_{j=l,r}\langle j|\hat{z}|j\rangle |j\rangle \langle j|$, and $\hat{I}$ is the corresponding identity operator. By convention, we set the center location of the double-well potential as the origin of coordinates (such that $ \langle l|\hat{z}|l\rangle =-\langle r|\hat{z}|r\rangle$), and drop the  additive physically-irreverent  constant $\langle |l|\hat{z}|l\rangle$, which leads the position operator to be simplified as $\hat{z}=|l\rangle \langle l|-|r\rangle \langle r| $. In the Wannier representation, $\varsigma =\hat{z}\otimes \hat{I}=\mathrm{diag}\left( 1,1,-1,-1 \right)
$, $\varLambda=\hat{z}\otimes \hat{\sigma}_z=\mathrm{diag}\left( 1,-1,-1,1 \right)$, such that we have\cite{citro2015spin},
 \begin{align}\label{con:14}
 P_a=&\left| c _{l\uparrow} \right |^2+\left| c _{l\downarrow} \right |^2-\left| c _{r\uparrow}\right |^2-\left| c _{r\downarrow} \right |^2,\nonumber\\
 P_s=&\left| c _{l\uparrow}\right |^2-\left| c _{l\downarrow}\right |^2-\left| c _{r\uparrow}\right |^2+\left| c _{r\downarrow} \right |^2.
 \end{align}
The
corresponding atomic density current ($I_a$) and the spin current ($I_s$) are given by\cite{zhang2012josephson, citro2015spin}
 \begin{equation}\label{con:15}
 I_a=\frac{d \langle \varPsi \left( t \right)|\varsigma|\varPsi \left( t \right) \rangle }{dt},
 I_s=\frac{d \langle \varPsi \left( t \right)|\varLambda|\varPsi \left( t \right) \rangle }{dt}.
 \end{equation}
According to the Schr\"{o}dinger equation and the definition of currents, the atomic current  and the spin current can be calculated as
\begin{align}\label{con:16}
 		I_a = &2\beta_l(\left| c _{l\uparrow}\right |^2+ \left| c _{l\downarrow}\right |^2)+2\beta_r(\left| c _{r\uparrow}\right |^2+ \left.| c _{r\downarrow}\right |^2) \nonumber\\&+2\nu \text {Im}(e^{-i\pi\gamma} {c _{l\downarrow}}c^{\ast} _{r\downarrow}-e^{i\pi\gamma} c _{r\downarrow}c^{\ast} _{l\downarrow}+e^{i\pi\gamma} {c _{l\uparrow}}c _{r\uparrow}^{\ast}-e^{-i\pi\gamma} c^{\ast}_{l\uparrow}c _{r\uparrow}),\nonumber\\
 		I_s=&2\beta_l(\left| c _{l\uparrow}\right |^2- \left| c _{l\downarrow}\right |^2)+2\beta_r(\left| c _{r\uparrow}\right |^2-\left| c _{r\downarrow}\right |^2)\nonumber\\&-2\nu \text {Im}[(e^{-i\pi\gamma} {c _{r\uparrow}}c^{\ast} _{l\uparrow}-e^{i\pi\gamma} c _{l\uparrow}c _{r\uparrow}^{\ast}+e^{-i\pi\gamma} {c _{l\downarrow}}c _{r\downarrow}^{\ast}-e^{i\pi\gamma} c _{r\downarrow}c _{l\downarrow}^{\ast})\nonumber\\&-\varOmega \text {Im}(c_{l\uparrow}c^{\ast} _{l\downarrow}-c _{l\uparrow}^{\ast} c_{l\downarrow}+c _{r\downarrow}c^{\ast} _{r\uparrow}-c _{r\uparrow}c^{\ast} _{r\downarrow})].
 	\end{align}

\begin{figure}[htbp]
\center
\includegraphics[width=8cm]{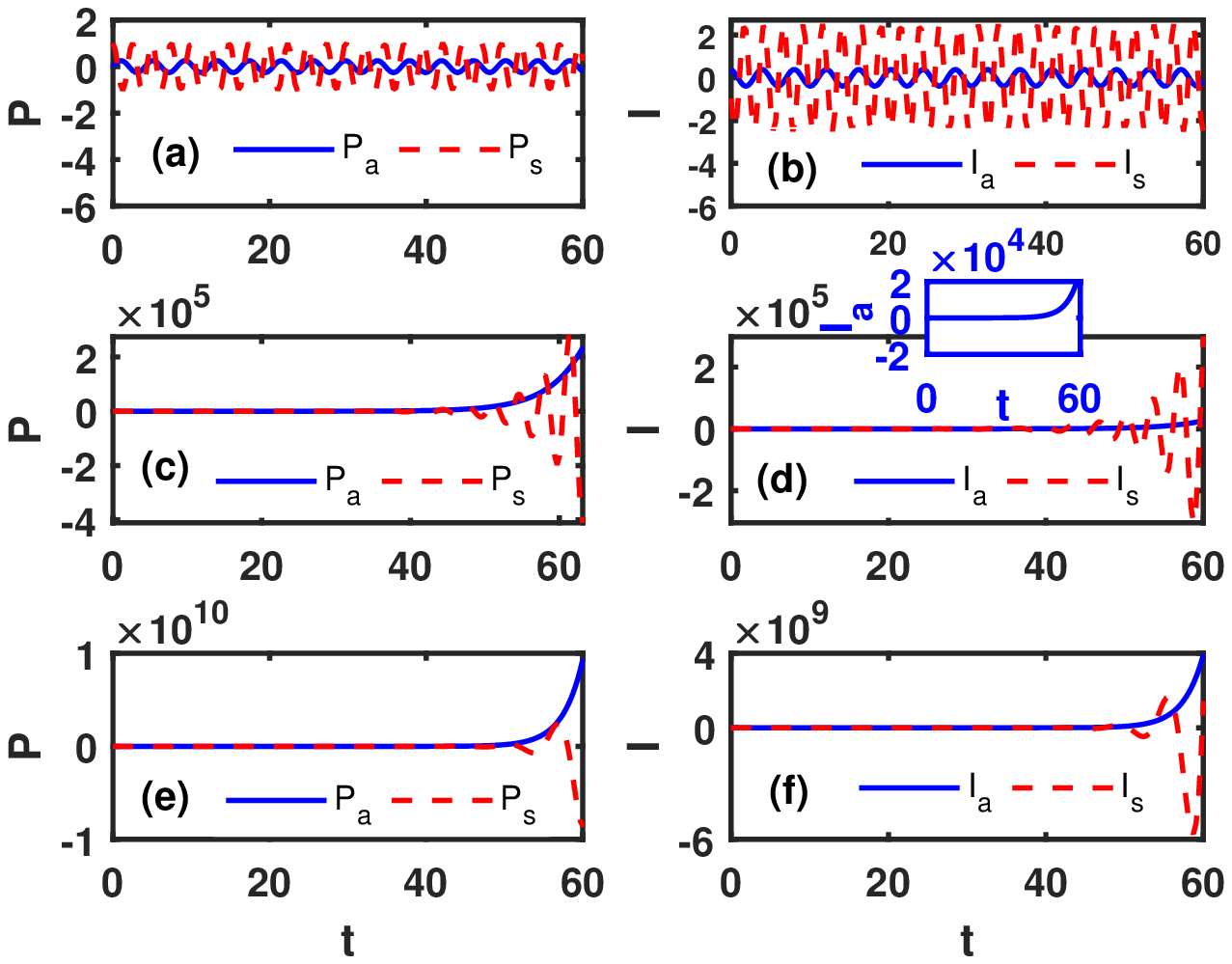}
\caption{(color online) Time-evolution curves of the population imbalance and magnetization (left column), and the atomic and spin currents (right column),  obtained from the $\mathcal{P} \mathcal{T}$-symmetric  system  \eqref{con:1} with the initial state prepared as $|\varPsi \left( 0 \right) \rangle=\left(c_{l\uparrow},~c_{l\downarrow},~c_{r\uparrow},~c_{r\downarrow} \right) ^{\textrm{T}}=\left( \frac{1}{\sqrt{2}},~0,~0,~\frac{1}{\sqrt{2}} \right) ^{\textrm{T}}$. The parameters are $\nu=2, \varOmega=1, \beta=0.2, \omega=20$, with (a)-(b) $\alpha=40, \gamma=0$; (c)-(d) $\alpha=40, \gamma=0.5$; (e)-(f) $\alpha=24, \gamma=0$. The blue solid line in the left column represents the population imbalance, and the red dashed line represents the magnetization; in the right column the blue solid line represents the atomic current, and the red dashed line represents the spin current.  In panel (d), we put an inset with  blue border for clear illustration of the exponential growth of the atomic current.}
\label{figure2}
\end{figure}

In figure \ref{figure2}, we exhibit the time evolutions of population imbalance, magnetization, spin current and atomic current  by numerically solving the time-dependent Schr\"{o}dinger
equation with Hamiltonian
 \eqref{con:1} for  fixed parameters $\nu=2, \varOmega=1, \beta=0.2, \omega=20$.
The initial state is taken as $|\varPsi \left( 0 \right) \rangle =\left( c_{l\uparrow},c_{l\downarrow},c_{r\uparrow},c_{r\downarrow} \right) ^{\textrm{T}}=\left( \frac{1}{\sqrt{2}},0,0,\frac{1}{\sqrt{2}} \right) ^{\textrm{T}}$. The blue solid line in the left (right) column represents population imbalance (atomic current), respectively, and the red dashed line in the left (right) column denotes magnetization (spin current), respectively. In figure \ref{figure2} (a)-(b), under the condition of $\alpha=40$ and $\gamma=0$, we reveal that both the population
imbalance (magnetization) and the atomic (spin) current exhibit periodic and stable oscillations, which implies that
the system is in the unbroken $\mathcal{PT}$ phase. When the driving amplitude is fixed and only the effective SOC strength is changed to $\gamma=0.5$, as shown in figure \ref{figure2} (c)-(d), we observe
that the population imbalance and atomic current increase exponentially without bound, while the magnetization and spin current oscillate up
and down around zero with oscillation amplitude increasing
continuously, indicating that the system enters into the broken $\mathcal{PT}$ phase.
In figure \ref{figure2} (d), we put an inset with blue border to clearly show the smooth  exponential
growth of the atomic current. Likewise, the transition from periodic (bounded) oscillation (unbroken $\mathcal{PT}$ phase) to  secular (unbounded) growth (broken $\mathcal{PT}$ phase) can be also achieved by tuning the driving parameters, as illustrated in figure \ref{figure2} (e)-(f), where we only change the
the driving amplitude to $\alpha=24$ and keep all the other parameters unchanged as compared to figure \ref{figure2} (a)-(b). The numerical results illustrate that apart from  periodic driving schemes, the management of SOC strength provides a flexible alternative to control the $\mathcal{PT}$ phase transition.

\begin{figure*}
 	\includegraphics[height=1.3in,width=2.2in]{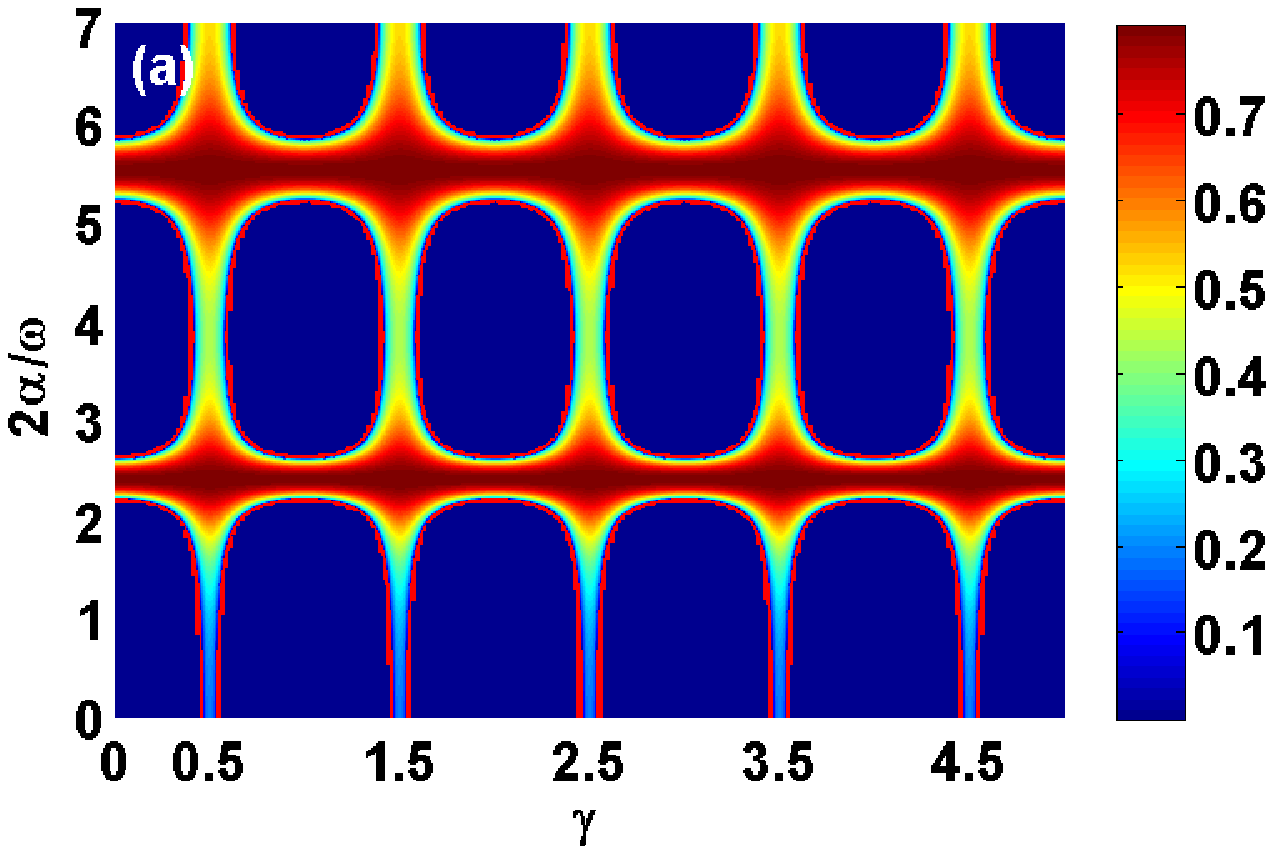}
 	\includegraphics[height=1.3in,width=2.2in]{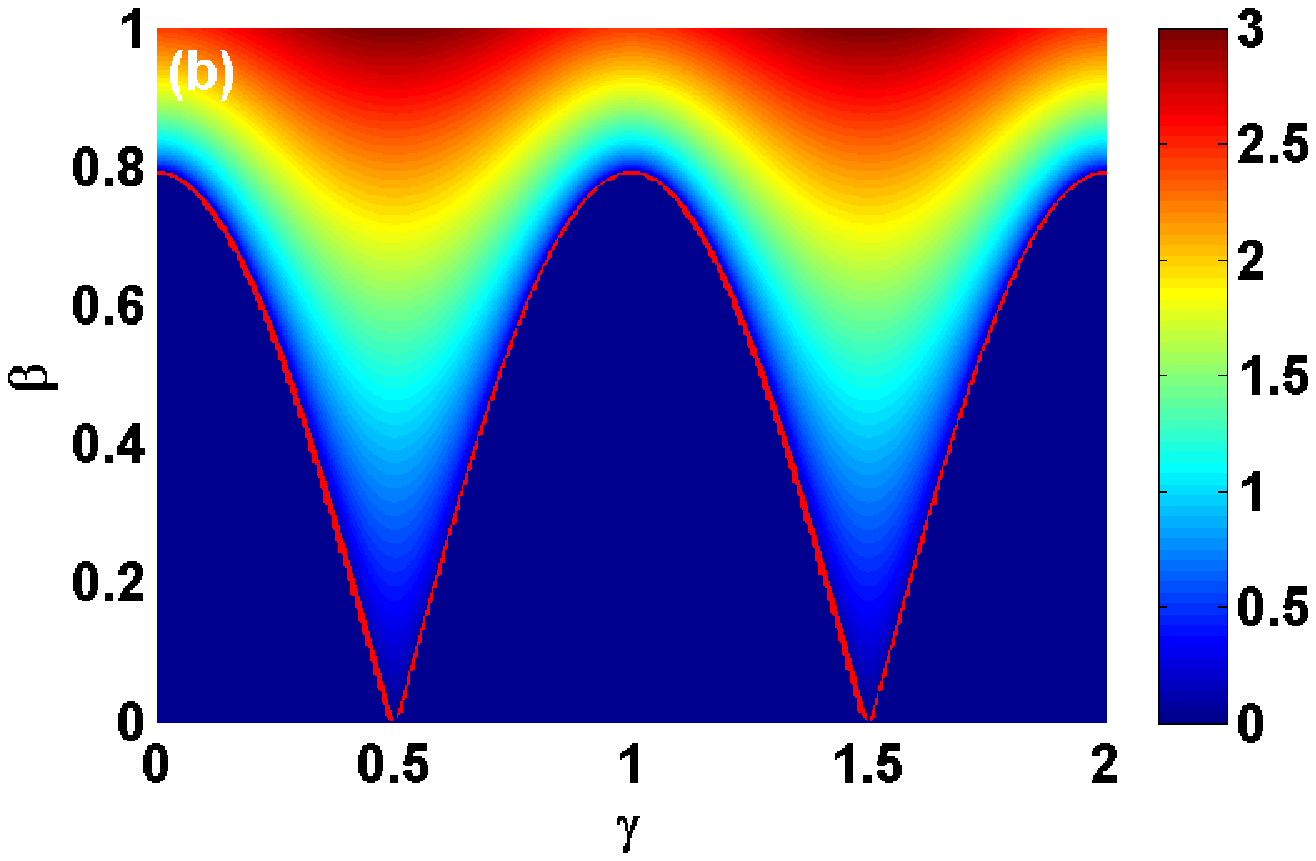}
 	\includegraphics[height=1.3in,width=2.2in]{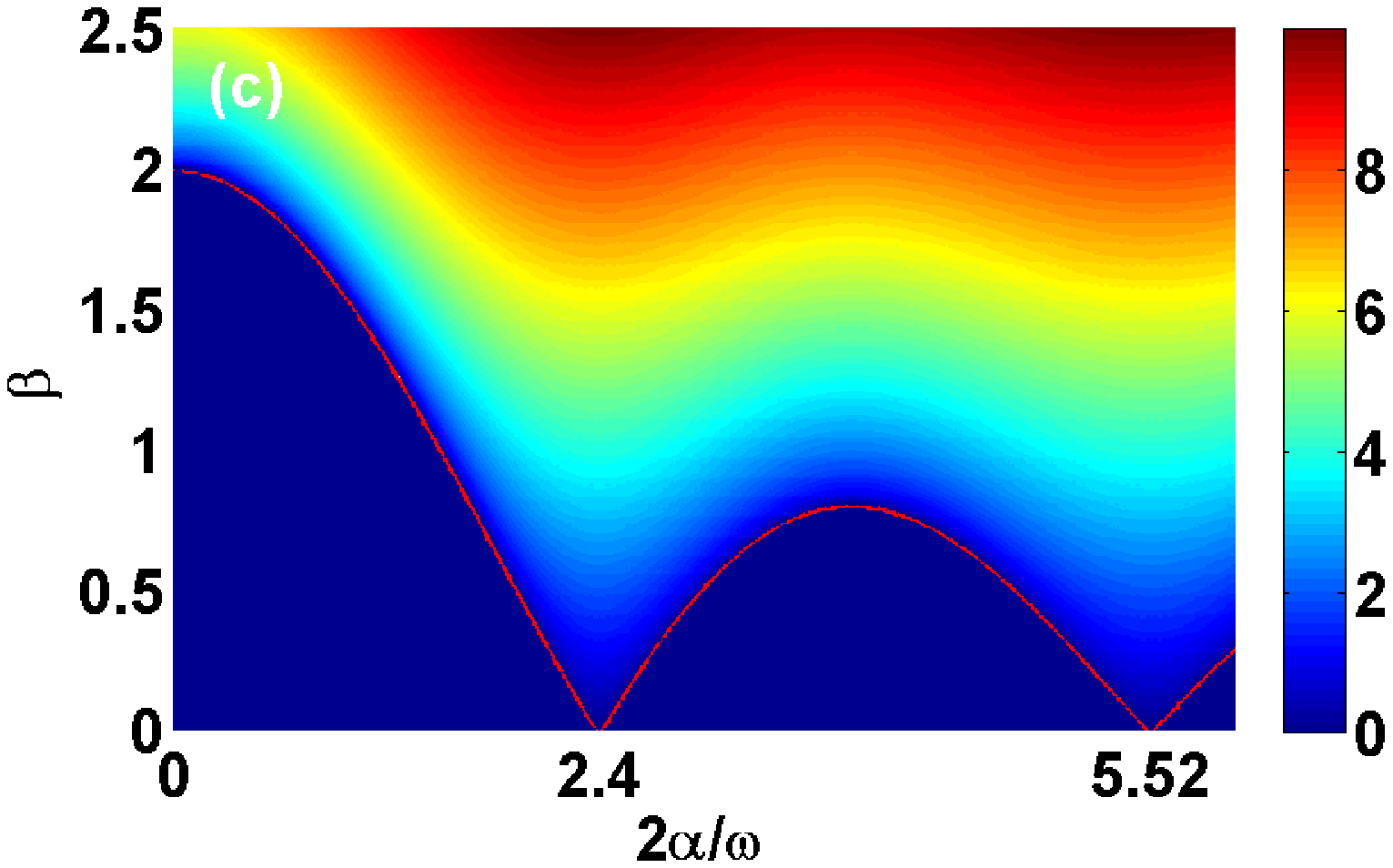}
 	\caption{Phase diagrams obtained by numerically calculating $|\text {Im}(E_2)|+ |\text {Im}(E_4)|$,  according to Eq.~\eqref{con:18}, in the parameter space (a) $(\gamma, 2\alpha/\omega)$ with $\beta=0.2$,
 (b) $(\gamma, \beta)$ with $2\alpha/\omega=4$, (c) $(2\alpha/\omega, \beta)$ with $\gamma=0$.
 The different map colors  specify different values of $|\text {Im}(E_2)|+ |\text {Im}(E_4)|$.
 The blue areas with $|\text {Im}(E_2)|+ |\text {Im}(E_4)|=0$
indicate the unbroken $\mathcal{PT}$-symmetric regions where the quasienergies are entirely real, and the red lines are the boundaries between the unbroken and broken $\mathcal{PT}$-symmetric regions, across which the quasienergies change from being all real to partial complex. Note that when $|\text {Im}(E_2)|+ |\text {Im}(E_4)|=0$, all of quasienergies are real.}
 \label{figure3}
 \end{figure*}

  Like in the undrvien system, the Floquet $\mathcal{PT}$-symmetric  system  \eqref{con:1}   will be said to be in the
unbroken $\mathcal{PT}$ phase whenever the quasienergies
are all real, whereas it is said to be in the broken $\mathcal{PT}$ phase if
complex conjugate quasienergies arise. The analytical expressions of quasienergies  \eqref{con:9} allow us to determine the accurate boundaries between the unbroken $\mathcal{PT}$ and broken $\mathcal{PT}$ phases.
From Eq.~\eqref{con:9}, under balanced gain and loss, the quasienergies become
\begin{align}\label{con:18}
&E_1=-E_2=-\frac{1}{2}\rho_1,~E_3=-E_4=-\frac{1}{2}\rho_2,~\nonumber\\&
\rho _1=\sqrt{m-w},~\rho _2=\sqrt{m+w},\nonumber\\&m=4\nu ^2\mathcal{J}_0 ^2\left( 2\alpha /\omega \right) +\varOmega ^2-4\beta ^2,\nonumber\\&
w=4\varOmega \sqrt{\left[ \nu \mathcal{J}_0 \left( 2\alpha /\omega \right) \cos \left( \gamma \pi \right) \right] ^2-\beta ^2}.
\end{align}

If the following two parameter relationships
\begin{align}\label{con:17}
&[\nu\mathcal{J} _0\left( {2\alpha}/{\omega} \right)]^2\cos ^2\left( \gamma \pi \right) \geqslant \beta ^2,\nonumber\\&4[\nu\mathcal{J} _0\left( {2\alpha}/{\omega} \right)]^2+\varOmega ^2- 4\beta ^2\geqslant 4\varOmega \sqrt{[\nu\mathcal{J} _0\left( {2\alpha}/{\omega} \right)]^2\cos ^2\left( \gamma \pi \right) -\beta ^2},
\end{align}
are satisfied,
we obtain four all-real quasienergies, then the system is in the
unbroken $\mathcal{PT}$ phase. Otherwise, if either one of the parametric relations in Eq.~\eqref{con:17} is not satisfied, at least two of the quasienergies will become complex, then the system is in the
broken $\mathcal{PT}$ phase. The ``=" signs taken in inequalities of \eqref{con:17} give the boundary (phase transition point) between unbroken  $\mathcal{PT}$-symmetric  and broken  $\mathcal{PT}$-symmetric regions. According to Eq.~\eqref{con:18}, we have drawn the  phase diagram  by numerically computing the values of $|\text {Im}(E_2)|+ |\text {Im}(E_4)|$, as shown in figure \ref{figure3}. By virtue of the relations $E_1=-E_2, E_3=-E_4$, we know that  if $|\text {Im}(E_2)|+ |\text {Im}(E_4)|=0$ holds, all quasienergies  naturally have no imaginary part. Thus, the blue areas with
$|\text {Im}(E_2)|+ |\text {Im}(E_4)|=0$  correspond to the unbroken  $\mathcal{PT}$-symmetric regions where the quasienergy spectrum is entirely real, and the areas with other colors correspond to the broken  $\mathcal{PT}$-symmetric regions. In figure \ref{figure3} (a), we set the parameter $\beta=0.2$ and plot $\left| \left. \text {Im}\left( E_2 \right) \right| \right. +\left| \left. \text {Im}\left( E_4 \right) \right| \right.$ as a function of $\gamma$ and $2\alpha/\omega$, where the red lines mark the boundary between unbroken  $\mathcal{PT}$-symmetric  and broken  $\mathcal{PT}$-symmetric regions. From figure \ref{figure3} (a), it is clearly seen that when either the driving parameters $2\alpha/\omega$ take the zeros of Bessel function such as $2\alpha/\omega=2.4, 5.52$..., or the effective  SOC is half-integer such as $\gamma=0.5, 1.5$..., the system is always in broken $\mathcal{PT}$ phase. The  $\mathcal{PT}$ phase transition
 can also be observed in the parameter space $(\gamma, \beta)$ with fixed $2\alpha/\omega=4$ and  the parameter space $(2\alpha/\omega, \beta)$ with  $\gamma=0$, as shown in figures \ref{figure3} (b) and (c)
 respectively. From these two plots, we further observe that when the effective SOC takes half-integer value or the driving parameters $2\alpha/\omega$ take the zeros of Bessel function, the $\mathcal{PT}$-symmetry-breaking occurs for arbitrarily small gain-loss coefficient. The salient features can be readily inferred from the  analytical expressions of quasienergies  \eqref{con:18}, where we find that when $\mathcal{J}_0 \left( 2\alpha /\omega \right)=0$ or $\cos \left( \gamma \pi \right)=0$, the term $w=4\varOmega \sqrt{\left[ \nu \mathcal{J}_0 \left( 2\alpha /\omega \right) \cos \left( \gamma \pi \right) \right] ^2-\beta ^2}$ becomes purely imaginary for arbitrarily small  $\beta$. This implies that
managing effective SOC alone can allow for spontaneous $\mathcal{PT}$-symmetry-breaking transition for arbitrary values of the gain and loss parameter, which has the same effect as the use of  periodic driving schemes.

As a next step we make further investigations on the current behaviors of the non-Hermitian system in both unbroken and broken $\mathcal{PT}$ phases.  In the
non-Hermitian system, the norm of the vector state (the quantum probability) $\mathcal{N}=\left< \left. \varPsi \right| \right. \left. \varPsi \right>=\Sigma_{j,\sigma}|c_{j,\sigma}|^2$ is not conserved with time evolution. To
eliminate the
contribution of the norm to the physical quantities, we  define the normalized  population imbalance and the magnetization as
\begin{align}\label{con:19}
P_{an}=\frac{P_a}{\mathcal{N}},P_{sn}=\frac{P_s}{\mathcal{N}},
\end{align}
and normalized atomic current and spin current,
\begin{align}\label{con:20}
I_{an}=\frac{dP_{an}}{dt}, I_{sn}=\frac{dP_{sn}}{dt}.
\end{align}
The above definition is more physically reasonable to reflect the population transfer and current behaviors. Especially, when the system is in the broken $\mathcal{PT}$ phase,  the norm $\mathcal{N}$ of quantum
states  will exponentially amplify with time due to the appearance
of complex quasienergies. Therefore, it is necessary to cancel the contribution
from the growth of the norm to the atomic population exchange and spin exchange between the two wells of the potential\cite{Valle2013, Luo2022}.

\begin{figure}[htbp]
\center
\includegraphics[width=8cm]{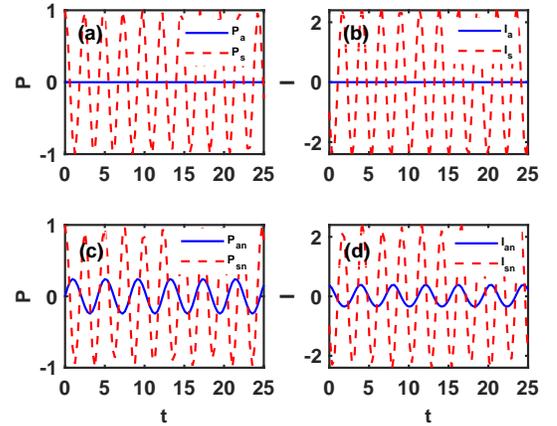}
\caption{Time-evolution curves of the population imbalance and magnetization (left column), and the atomic and spin currents (right column),  obtained from the original system  \eqref{con:1} with the initial state prepared as $|\varPsi \left( 0 \right) \rangle=\left(c_{l\uparrow},~c_{l\downarrow},~c_{r\uparrow},~c_{r\downarrow} \right) ^{\textrm{T}}=\left( \frac{1}{\sqrt{2}},~0,~0,~\frac{1}{\sqrt{2}} \right) ^{\textrm{T}}$. The parameters are $\nu=2,~\varOmega=1,~\gamma=0,~\omega=20,~\alpha=40$, with (a)-(b) $\beta=0$ and (c)-(d) $\beta=0.2$. In panels (c) and (d), all physical quantities
 have been normalized.}
\label{figure4}
\end{figure}
As reported in Ref.~\cite{zhang2012josephson}, a net  spin current (i.e., the spin current is nonzero while atomic current is zero) can be induced in spin-orbit-coupled BJJ for weak Raman coupling. In the Hermitian case, our numerical investigation reveals that such a  net  spin current
(together with a vanishing atomic current) can be observable for arbitrary values of  Raman coupling strength, when the system is initialized in state $|\varPsi \left( 0 \right) \rangle =\left( c_{l\uparrow},~c_{l\downarrow},~c_{r\uparrow},~c_{r\downarrow} \right) ^{\textrm{T}}=\left( \frac{1}{\sqrt{2}},~0,~0,~\frac{1}{\sqrt{2}} \right) ^{\textrm{T}}$. As an example, we take the parameter set,
$\beta _l=\beta _r=0, \nu=2,~\varOmega=1,~\gamma=0,~\omega=20,~\alpha=40$, and from the time-dependent Schr\"{o}dinger
equation with Hamiltonian \eqref{con:1}, we plot the time evolution of population imbalance (magnetization) and the atomic (spin) current given by the formula \eqref{con:16} in figures \ref{figure4} (a) and (b) respectively. As expected, figures \ref{figure4} (a) and (b) show that the magnetization and spin current exhibit periodic oscillation, while the population imbalance and atomic current are always zero.
When introducing the balanced gain and loss term $\beta _l=\beta _r=0.2$ and keeping the other parameters fixed, the system is in the unbroken $\mathcal{PT}$ phase, and figures \ref{figure4} (c) and (d) show that all normalized physical quantities such as the population imbalance (magnetization) and the atomic (spin) current exhibit periodic oscillations, which indicates that the net  spin current is destroyed by  the non-Hermiticity in the unbroken $\mathcal{PT}$-symmetric  region.

Now we move on to investigate the current behaviors in the broken $\mathcal{PT}$-symmetric  region.
When the non-Hermiticity degree $\beta$ is sufficiently strong, the term $w$ in Eq.~\eqref{con:18} becomes a purely imaginary number, $w=iw'=i4\varOmega \sqrt{\beta ^2-\left[ \nu \mathcal{J}_0 \left( 2\alpha /\omega \right) \cos \left( \gamma \pi \right) \right] ^2}$.
According to Eq.~\eqref{con:18}, we get
\begin{align}\label{con:22}
&E_{1,2}=\mp \frac{1}{2}\sqrt{m-iw^{'}}=\mp \frac{1}{2}\left( m^2+w^{'2} \right) ^{\frac{1}{4}}\left( \cos \frac{\phi ^{'}}{2}+i\sin \frac{\phi ^{'}}{2} \right),\nonumber\\&
E_{3,4}=\mp \frac{1}{2}\sqrt{m+iw^{'}}=\mp \frac{1}{2}\left( m^2+w^{'2} \right) ^{\frac{1}{4}}\left( -\cos \frac{\phi ^{'}}{2}+i\sin \frac{\phi ^{'}}{2} \right),
\end{align}
where we have defined $m-iw^{'}=\left( m^2+w^{'2} \right) ^{\frac{1}{2}}e^{i\phi ^{'}}$ and
$m+iw^{'}=\left( m^2+w^{'2} \right) ^{\frac{1}{2}}e^{i(2\pi-\phi ^{'})}$.
For sufficiently strong gain-loss coefficient $\beta$, we have $m<0, w^{'}>0$ and $\phi ^{'}\in \left( \pi ,\frac{3\pi}{2} \right)$, such that
\begin{align}\label{con:23}
&\mathrm{Im}\left( E_2 \right) =\mathrm{Im}\left( E_4 \right) =\frac{1}{2}\left( m^2+w^{'2} \right) ^{\frac{1}{4}}\sin \frac{\phi ^{'}}{2}>0,\nonumber\\&
\mathrm{Im}\left( E_1 \right) =\mathrm{Im}\left( E_3 \right) =-\frac{1}{2}\left( m^2+w^{'2} \right) ^{\frac{1}{4}}\sin \frac{\phi ^{'}}{2}<0.
\end{align}
That is to say, the imaginary parts of quasienergies $E_1$ and $E_3$ in Eq.~\eqref{con:18} are negative, and the imaginary parts of quasienergies $E_2$ and $E_4$ are positive.

To gain analytical insight into the  current behaviors in the broken $\mathcal{PT}$-symmetric  region,
we expand the quantum state at the initial time in the basis of Floquet modes, i.e., $
\left| \left. \varPsi \left( 0 \right) \right> \right. =\sum_{n=1}^4{a_n}\left| \left. \varphi _n(0) \right> \right.= \sum_{n=1}^4{a_n\left| \left. \varphi _{n}^{'} \right> \right.}$.
At time $t$, the wave function evolves according to Eq.~\eqref{evqutumstate}. As time increases, the components $a_n$  with $\textrm{Im}(E_n)>0$ will exponentially
grow, and that of negative $\textrm{Im}(E_n)$ exponentially decays.
Thus, the asymptotic solution of the time-evolved quantum state can be written as
\begin{align}\label{con:24}
\left| \left. \varPsi \left( t\rightarrow \infty \right) \right> \right.&=a_2\left| \left. \varphi _2 (t)\right> \right. e^{-i\varepsilon_2t} +a_4\left| \left.\varphi _4 (t)\right> \right. e^{-i\varepsilon_4t}\nonumber\\&= e^{\mathrm{Im}\left( E_2 \right) t}\left( a_2\hat{S}\left| \left. \varphi _{2}^{'} \right> e^{-i\mathrm{Re}\left( E_2 \right) t}+a_4\hat{S}\left| \left. \varphi _{4}^{'} \right> \right. e^{-i\mathrm{Re}\left( E_4 \right) t} \right. \right),
\end{align}
where $\hat{S}=\mathrm{diag}\left( e^{-i\frac{\alpha}{\omega}\sin \left( \omega t \right)},e^{-i\frac{\alpha}{\omega}\sin \left( \omega t \right)},e^{i\frac{\alpha}{\omega}\sin \left( \omega t \right)},e^{i\frac{\alpha}{\omega}\sin \left( \omega t \right)} \right)$.

From Eq.~\eqref{con:24}, we have the asymptotic norm
\begin{align}\label{con:26}
\mathcal{N} \left( t\rightarrow \infty \right) &=\left< \left. \varPsi(t\rightarrow \infty ) \right| \right. \left. \varPsi(t\rightarrow \infty ) \right>
\nonumber\\&=e^{2\mathrm{Im}\left( E_2 \right) t}\left( \left| a_2 \right|^2\left< \left. \varphi _{2}^{'} \right| \right. \left. \varphi _{2}^{'} \right> +\left| a_4 \right|^2\left< \left. \varphi _{4}^{'} \right| \right. \left. \varphi _{4}^{'} \right> \right) \nonumber\\&+e^{2\mathrm{Im}(E_2)t}\left(a_{2}^{\ast}a_4\left< \left. \varphi _{2}^{'} \right| \left. \varphi _{4}^{'} \right> \right. e^{i\mathrm{Re}\left( E_2-E_4 \right) t}+H.c.\right),
\end{align}
and the asymptotic population imbalance and magnetization
\begin{align}\label{con:27}
P_a\left( t\rightarrow \infty \right)&=\left< \left. \varPsi (t\rightarrow \infty ) \right| \right.\varsigma\left| \left. \varPsi (t\rightarrow \infty )  \right> \right. \nonumber\\&=e^{2\mathrm{Im}\left( E_2 \right) t}\left( \left| a_2 \right|^2\left< \left. \varphi _{2}^{'} \right| \right. \varsigma\left| \left. \varphi _{2}^{'} \right> + \right. \left| a_4 \right|^2\left< \left. \varphi _{4}^{'} \right| \right.\varsigma\left| \left. \varphi _{4}^{'} \right> \right. \right)\nonumber\\&+ e^{2\mathrm{Im} (E_2)t}\left(a_{2}^{\ast}a_4\left< \left. \varphi _{2}^{'} \right| \right. \varsigma\left| \left. \varphi _{4}^{'} \right> \right. e^{i\mathrm{Re}\left( E_2-E_4 \right) t}+H.c.\right),
\end{align}
\begin{align}\label{con:28}
P_s\left( t\rightarrow \infty \right)&=\left< \left. \varPsi (t\rightarrow \infty ) \right| \right.\varLambda\left| \left. \varPsi (t\rightarrow \infty )\right> \right.\nonumber\\& =e^{2\mathrm{Im}\left( E_2 \right) t}\left( \left| a_2 \right|^2\left< \left. \varphi _{2}^{'} \right| \right. \varLambda\left| \left. \varphi _{2}^{'} \right> + \right. \left| a_4 \right|^2\left< \left. \varphi _{4}^{'} \right| \right. \varLambda\left| \left. \varphi _{4}^{'} \right> \right. \right)
\nonumber\\&+e^{2\mathrm{Im}(E_2) t}\left(a_{2}^{\ast}a_4\left< \left. \varphi _{2}^{'} \right| \right. \varLambda\left| \left. \varphi _{4}^{'} \right> \right. e^{i\mathrm{Re}\left( E_2-E_4 \right) t}+H.c.\right).
\end{align}

In principle, we can explicitly derive the asymptotic forms of all the physical quantities by inserting
the eigenvectors $|\varphi _n'\rangle$ of the effective Hamiltonian  \eqref{Heff} into Eqs.~\eqref{con:26},
\eqref{con:27} and \eqref{con:28}. Nevertheless, the derivation process is rather tedious and lengthy for general effective SOC strength. For simplicity, we take $\gamma =0$ as an example for illustration.
When $\gamma =0$, the quasienergies takes the simple form
\begin{align}\label{con:29}
&E_{1,2}=\mp \left( -\frac{\varOmega}{2}+i\sqrt{\beta ^2-\left[ \nu \mathcal{J} _0\left( \frac{2\alpha}{\omega} \right) \right] ^2} \right),\nonumber\\&
E_{3,4}=\mp \left( \frac{\varOmega}{2}+i\sqrt{\beta ^2-\left[ \nu \mathcal{J} _0\left( \frac{2\alpha}{\omega} \right) \right] ^2} \right),
\end{align}
with the corresponding eigenstates $
\left| \left. \varphi _{2}^{'} \right> \right.=\left( \begin{matrix}
	\begin{matrix}
	\zeta,&\zeta\\
\end{matrix},&		1,&		1\\
\end{matrix} \right) ^{\textrm{T}}
$,~$
\left| \left. \varphi _{4}^{'} \right> \right.=\left( -\begin{matrix}
	\begin{matrix}
	\zeta,&		\zeta\\
\end{matrix},&		-1,&		1\\
\end{matrix} \right) ^{\textrm{T}}
$,~where $
\zeta =i\left( \frac{\beta +\sqrt{\beta ^2-\nu ^2\mathcal{J} _{0}^{2}\left( \frac{2\alpha}{\omega} \right)}}{\nu \mathcal{J} _0\left( \frac{2\alpha}{\omega} \right)} \right)
$. Armed with these, we immediately have
\begin{align}
\mathcal{N} \left( t\rightarrow \infty \right) =2\left( \left| a_2 \right|^2+\left| a_4 \right|^2 \right) \left( \left| \zeta \right|^2+1 \right) e^{2\mathrm{Im}\left( E_2 \right) t},
\end{align}
\begin{align}\label{con:30}
P_a\left( t\rightarrow \infty \right)=2\left( \left| a_2 \right|^2+\left| a_4 \right|^2 \right) \left( \left| \zeta \right|^2-1 \right) e^{2\mathrm{Im}\left( E_2 \right) t},
\end{align}
\begin{align}
P_s\left( t\rightarrow \infty \right)=2\left( 1-\left| \zeta \right|^2 \right) \left( a_{2}^{\ast}a_4e^{i\varOmega t}+a_{4}^{\ast}a_2e^{-i\varOmega t} \right) e^{2\mathrm{Im}\left( E_2 \right) t}.
\end{align}
According to  Eq. \eqref{con:19}, we obtain
\begin{align}\label{con:31}
P_{an}\left( t\rightarrow \infty \right)=\frac{P_a\left( t\rightarrow \infty \right)}{\mathcal{N}}=\frac{\left| \zeta \right|^2-1}{\left| \zeta \right|^2+1},
\end{align}
\begin{align}\label{con:32}
P_{sn}\left( t\rightarrow \infty \right)=\frac{P_s\left( t\rightarrow \infty \right)}{\mathcal{N}}=\frac{\left( 1-\left| \zeta \right|^2\right) \left( a_{2}^{\ast}a_4e^{i\varOmega t}+a_{4}^{\ast}a_2e^{-i\varOmega t} \right)}{\left( \left| \zeta \right|^2+1 \right) \left( \left| a_2 \right|^2+\left| a_4 \right|^2 \right)}.
\end{align}
From the definition for normalized atomic current and spin current, we get
\begin{align}\label{con:33}
I_{an}=\frac{dP_{an}}{dt}\left( t\rightarrow \infty \right)=0.
\end{align}
\begin{align}\label{con:34}
I_{sn}=\frac{dP_{sn}}{dt}\left( t\rightarrow \infty \right)=\frac{2\varOmega \left( \left| \zeta \right|^2-1 \right) \mathrm{Im}\left( a_{2}^{\ast}a_4e^{i\varOmega t} \right)}{\left( \left| \zeta \right|^2+1 \right) \left( \left| a_2 \right|^2+\left| a_4 \right|^2 \right)}.
\end{align}
\noindent
\begin{figure}[htbp]
\center
\includegraphics[width=8cm]{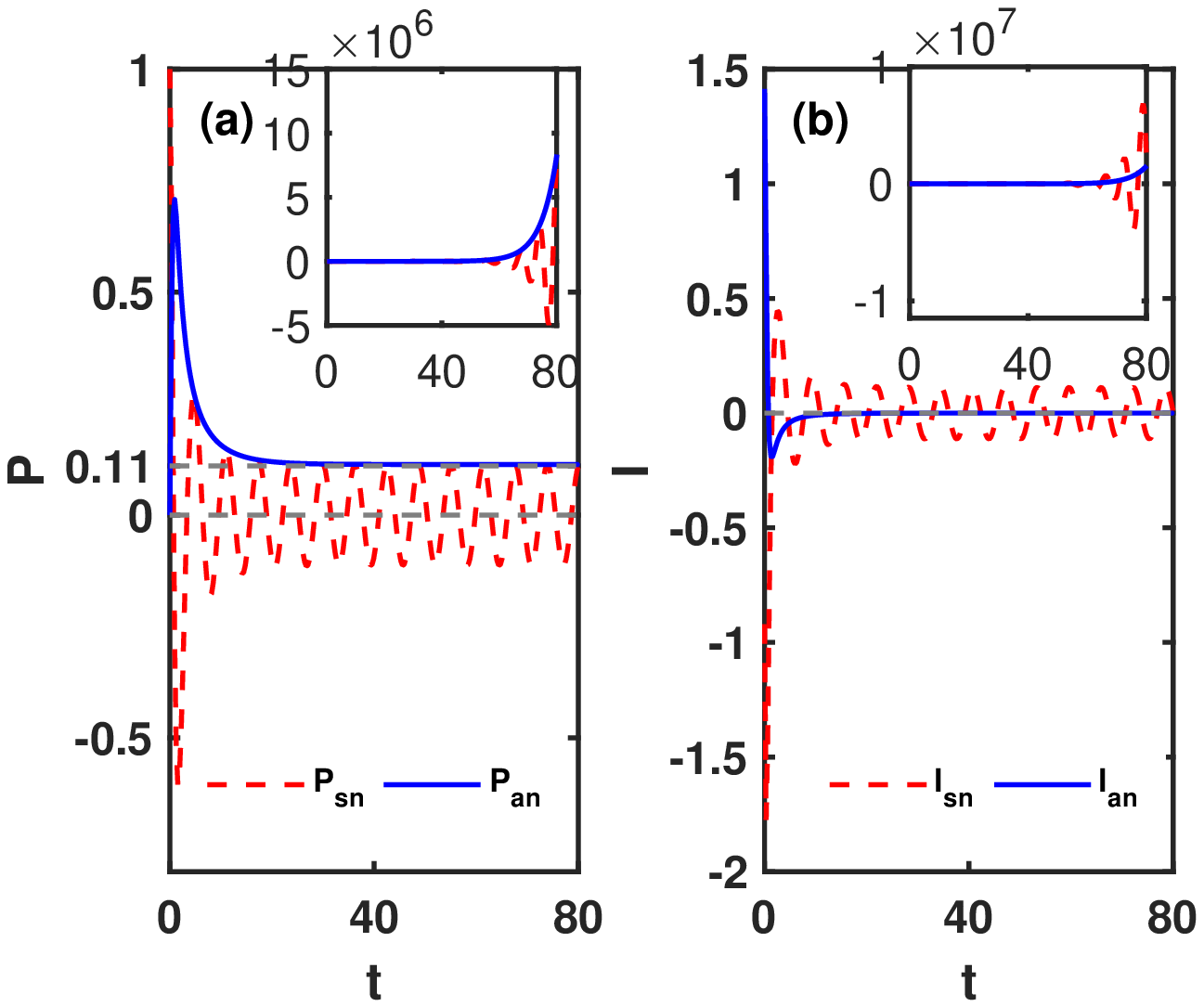}
\includegraphics[width=8cm]{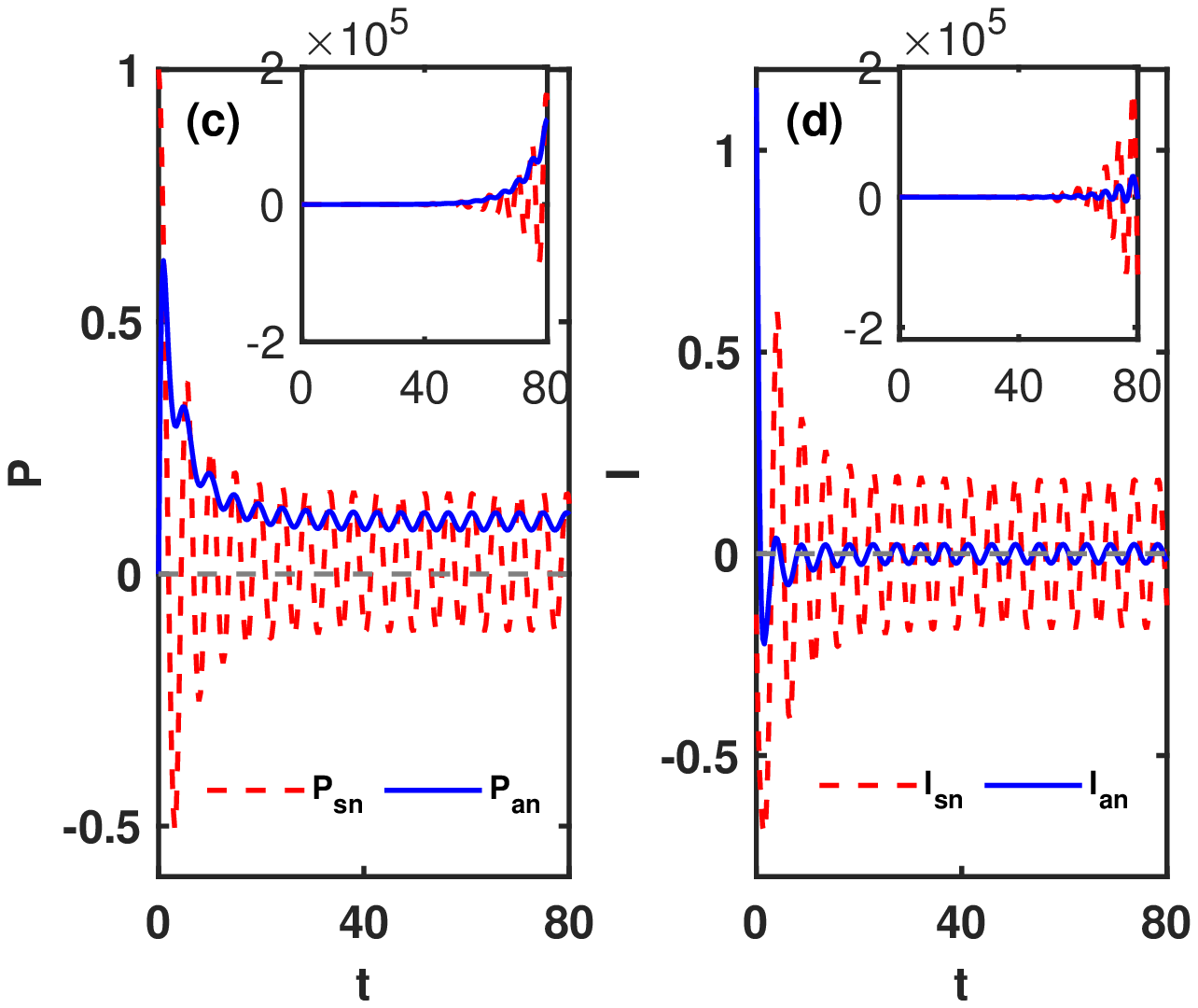}
\caption{Time-evolution curves of the  population imbalance and  magnetization (left column), and the  atomic and spin currents (right column),  obtained from the original system  \eqref{con:1}. The parameters are set as $\nu=2,~\varOmega=1,~\omega=20,~\alpha=40$, with  (a)-(b) $\beta=0.8,~\gamma=0$ and (c)-(d) $\beta=0.65,~\gamma=0.8$. The initial state is given by $|\varPsi \left( 0 \right) \rangle=\left(c_{l\uparrow},~c_{l\downarrow},~c_{r\uparrow},~c_{r\downarrow} \right) ^{\textrm{T}}=\left( \frac{1}{\sqrt{2}},~0,~0,~\frac{1}{\sqrt{2}} \right) ^{\textrm{T}}$.
The main panels show the
time evolutions of the normalized physical quantities, and the
insets in each panel give the time evolutions of the counterparts that are not normalized.}
\label{figure5}
\end{figure}

Eqs.~\eqref{con:33} and \eqref{con:34} mean that if we drop the contribution of the exponential growth of the norm to the current in the  broken $\mathcal{PT}$-symmetric region,  for the case of $\gamma=0$, the spin current is nonzero while the atomic current is zero. Along the very same line of  reasoning, for the broken $\mathcal{PT}$-symmetric phase, we can analytically demonstrate  that in the specific case of $\gamma=0.5$ (or half-integer values of $\gamma$), the situation is the same as in $\gamma=0$, where the normalized atomic current is zero and normalized spin current is not zero.
As we know, the normalized current is more appropriate to quantify the non-Hermitian physics than its unnormalized counterpart. In this sense, we can say that the net spin current (zero $I_{an}$ and nonzero $I_{sn}$) exists in the broken $\mathcal{PT}$-symmetric region, which we find is nevertheless not a general feature for arbitrary values of effective  SOC strength. The above conclusions are independent of the preparation of initial state.

 In figure \ref{figure5}, we show the time-evolution curves of the normalized physical quantities such as the population imbalance (magnetization) and the atomic (spin) current, on the basis of full numerical analysis of the system \eqref{con:1} with the same initial state as before. The system parameters are fixed as
 $\nu=2,~\varOmega=1,~\omega=20,~\alpha=40$, with (a)-(b) $\beta=0.8,~\gamma=0$ and (c)-(d) $\beta=0.65,~\gamma=0.8$. The main figures show the time evolutions of the normalized physical quantities, and the insets in each panel give the time evolutions of the unnormalized counterparts. As illustrated in the two insets of the left column,
  the unnormalized population imbalances (blue lines) for both parameter sets  show unbounded growth as a signature of the broken $\mathcal{PT}$ phase. The difference is that the unnormalized  population imbalance  shows smooth exponential behavior for the case of $\gamma=0$, while for $\gamma=0.8$, due to the general nature that the complex Floquet eigenstates are not necessarily orthogonal in non-Hermitian  system,
  the unnormalized  population imbalance shows oscillatory growth (exponential growth plus periodic oscillation), rather than smooth exponential growth.
  As shown in figure \ref{figure5} (a)-(b) with $\beta=0.8$ and $\gamma=0$, the normalized  population imbalance  oscillates during the initial
short time interval, after which it is asymptotically unchanged
with time evolution, and as the time evolves, the vanishing normalized atomic current will be a consequence [see the blue line in main figure of panel (b)]. At the same time, the normalized magnetization  and normalized spin current (see red dashed lines) show stable periodic oscillations after a transient decay.
When  $\beta=0.65$ and $\gamma=0.8$ are set as shown in figure \ref{figure5} (c)-(d), the normalized  population imbalance rapidly tends to a small-amplitude periodic oscillation around a nonzero value, and as a result,
the normalized atomic current  asymptotically tends to a stable periodic oscillation around zero with very small oscillating amplitude. Additionally,  the normalized magnetization  and normalized spin current exhibit stable periodic oscillations with large amplitude as the increase in time. As the numerical simulations demonstrate, the net spin current (zero $I_{an}$ and nonzero $I_{sn}$) exists in the broken $\mathcal{PT}$-symmetric region, but it is truly not a general character for arbitrary values (as exemplified by $\gamma=0.8$) of effective SOC strength. The numerical results are well consistent with the analytical ones. For example, given the initial condition and system parameters as in figure \ref{figure5} (a), from Eqs.~\eqref{con:31}-\eqref{con:34} we can explicitly determine the  asymptotical solutions of  all normalized physical quantities such as the population imbalance (magnetization) and the atomic (spin) current. We mark  $P_{an}=0.11$ obtained from Eq.~\eqref{con:31} as
 horizontal line in figure \ref{figure5} (a), which is in perfect agreement with the numerical result based on the original system \eqref{con:1} and thus fully confirm the validity of the time-averaging method used in our work.

We remark that the  net spin current (zero $I_{an}$ and nonzero $I_{sn}$) effect in broken $\mathcal{PT}$-symmetric region has simple physical picture. We consider a  noninteracting BEC in a double well. When the stable symmetry-breaking dynamics is reached, the two spin components move in opposite directions, and there is no net-particle tunneling (that is, the number of the spin $\uparrow$ ($\downarrow$) atoms tunneling from the left to right well is equal to that of  the spin $\downarrow$ ($\uparrow$) atoms tunneling from the right to  left well). This is the same as the Hermitian system; the only exception is that  the atomic populations in two wells experience the same exponential-law growth, which effect can be removed by normalization.
 \section{Currents under unbalanced gain and loss}
In this section, we turn to explore the current behaviors  of the non-$\mathcal{PT}$-symmetric system with unbalanced gain and loss ($\beta_l\ne \beta_r$). We assume $\beta_l<\beta_r$, which represents  a dissipative system with the particle loss in the right well greater than the gain in the left well. Generally, in such a situation, all of $\textrm{Im}(E_n)$ are less than zero and the atomic probabilities will asymptotically decay to zero.
However, if we tune the driving parameters to
satisfy the following conditions:
\begin{align}\label{con:35}
4\mathrm[\nu\mathcal{J} _0\left({2\alpha}/{\omega} \right)]^2+\varOmega ^2=\left( \beta_l+\beta_r \right) ^2,~\beta_l-\beta_r=-\sqrt{\varOmega x},
\end{align}
where $x=\sqrt{(\beta_l+\beta_r)^2-4(\nu\mathcal{J} _0\left( {2\alpha}/{\omega} \right))^2\cos^2\left(\gamma \pi\right)}$, from  Eq. \eqref{con:9} the quasienergies are given by
\begin{align}\label{con:36}
&E_1=i\left( \beta _l-\beta _r \right)-\frac{1}{2}\sqrt{\varOmega x},~E_2=\frac{1}{2}\sqrt{\varOmega x},~\nonumber\\&E_3=i\left( \beta _l-\beta _r \right) +\frac{1}{2}\sqrt{\varOmega x},~E_4=-\frac{1}{2}\sqrt{\varOmega x}.
\end{align}
  That is to say, two of the quasienergies are real, and the imaginary parts of the other two are less than zero. This will result in that the populations and currents tend to be stable as time increases. According to  Eq.~\eqref{evqutumstate}, at $t\rightarrow \infty$, the asymptotic solution of the quantum state can be written as
\begin{align}\label{con:37}
\left| \left. \varPsi \left( t\rightarrow \infty \right) \right> \right.&=a_2\left| \left. \varphi _2 (t)\right> \right. e^{-i\varepsilon_2t}+ a_4\left| \left.\varphi _4 (t)\right> \right. e^{-i\varepsilon_4t}\nonumber\\&= a_2\hat{S}\left| \left.  \varphi _{2}^{'} \right> e^{-iE_2t} \right. +a_4\hat{S}\left| \left.  \varphi _{4}^{'} \right> e^{-iE_4t} \right.,
\end{align}
where $\hat{S}=\mathrm{diag}\left( e^{-i\frac{\alpha}{\omega}\sin \left( \omega t \right)},e^{-i\frac{\alpha}{\omega}\sin \left( \omega t \right)},e^{i\frac{\alpha}{\omega}\sin \left( \omega t \right)},e^{i\frac{\alpha}{\omega}\sin \left( \omega t \right)} \right)$.
In our analysis, we focus on the case $\gamma=0.5$,
but similar behaviors can be obtained for other choices of effective SOC as well. When $\gamma=0.5$, from Eq.~\eqref{con:36} we can simplify the two real quasienergies as $E_2=\frac{\sqrt{\varOmega \left( \beta _l+\beta _r \right)}}{2}$, $E_4=-\frac{\sqrt{\varOmega \left( \beta _l+\beta _r \right)}}{2}$, with the corresponding  eigenstates $
\left| \left. \varphi _{2}^{'} \right> \right. =\left( \begin{matrix}
	\begin{matrix}
	\xi,&		\xi\\
\end{matrix},&		1,&		1\\
\end{matrix} \right) ^{\textrm{T}}
$,~$
\left| \left. \varphi _{4}^{'} \right> \right. =\left( \begin{matrix}
	\begin{matrix}
	-\xi ^{\ast},&		\xi ^{\ast}\\
\end{matrix},&		-1,&		1\\
\end{matrix} \right) ^{\textrm{T}}
$,  where $\xi =\frac{2\beta _r-i\varOmega -i\sqrt{\varOmega \left( \beta _l+\beta _r \right)}}{2\nu \mathcal{J} _0\left( 2\alpha /\omega \right)}$. Applying them to Eq.~\eqref{con:37} yields
\begin{align}\label{con:39}
&c_{l\uparrow,\downarrow}\left( t\rightarrow \infty \right)=[a_2\xi e^{-i\frac{\sqrt{\varOmega \left( \beta _l+\beta _r \right)}}{2}t}\mp a_4\xi ^{\ast}e^{i\frac{\sqrt{\varOmega \left( \beta _l+\beta _r \right)}}{2}t}]\times e^{-i\frac{\alpha}{\omega}\sin \left( \omega t \right)},\nonumber\\&
c_{r\uparrow ,\downarrow}\left( t\rightarrow \infty \right)=[\mp a_2e^{-i\frac{\sqrt{\varOmega \left( \beta _l+\beta _r \right)}}{2}t}+a_4e^{i\frac{\sqrt{\varOmega \left( \beta _l+\beta _r \right)}}{2}t}]\times e^{i\frac{\alpha}{\omega}\sin \left( \omega t \right)}.
\end{align}
According to the definition for population imbalance and  magnetization, we obtain
\begin{align}\label{con:40}
P_a\left( t\rightarrow \infty \right)&=\left( \left| c _{l\uparrow} \right |^2+\left| c _{l\downarrow} \right |^2-\left| c _{r\uparrow}\right |^2-\left| c _{r\downarrow} \right |^2\right) |_{t\rightarrow \infty }\nonumber\\&=
2\left( \mid \xi \mid ^2-1 \right) \left( \mid a_2\mid ^2+\mid a_4\mid ^2 \right),\nonumber\\
P_s\left( t\rightarrow \infty \right)&=\left(\left| c _{l\uparrow}\right |^2-\left| c _{l\downarrow}\right |^2-\left| c _{r\uparrow}\right |^2+\left| c _{r\downarrow} \right |^2\right) \mid_{t\rightarrow \infty }\nonumber\\&=
4\mid a_2\mid \mid a_4\mid  \left( \mid \xi \mid ^2+1 \right)\cos \left[ \theta _4-\theta _2+ \sqrt{\varOmega \left( \beta _l+\beta _r \right)}t \right],
\end{align}
where $a_2=|a_2|e^{i\theta _2},~a_4=|a_4|e^{i\theta _4}$. Substituting Eq. \eqref{con:39} into Eq. \eqref{con:16}, we get the atomic current and spin current as
\begin{align}\label{con:41}
I_a\left( t\rightarrow \infty \right)=&0,\nonumber\\
I_s\left( t\rightarrow \infty \right)=&-4\sqrt{\varOmega \left( \beta _l+\beta _r \right)}\mid a_2\mid \mid a_4\mid \left( \mid \xi \mid ^2+1 \right)\nonumber\\&
\times\sin \left[ \theta _4-\theta _2+ \sqrt{\varOmega \left( \beta _l+\beta _r \right)} t\right].
\end{align}
\begin{figure}[htbp]
\center
\includegraphics[width=8cm]{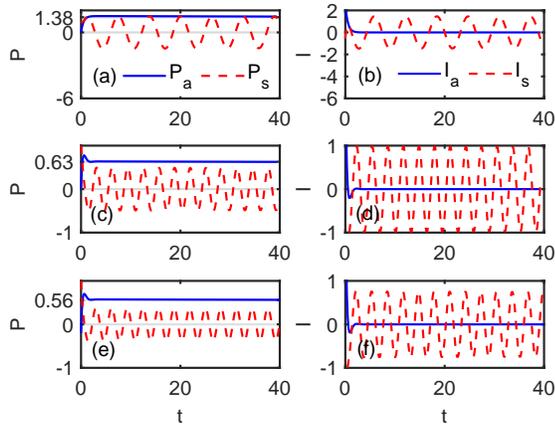}
\caption{Time-evolution curves of the population imbalance and magnetization (left column), and the atomic and spin currents (right column)  obtained from the non-$\mathcal{PT}$-symmetric system  \eqref{con:1} under the condition \eqref{con:35}. The parameters are set as (a)-(b) $\nu=2,~\varOmega=1,~\alpha=22.52,~\omega=40,~\beta_l=1,~\beta_r=2,~\gamma=0$; (c)-(d) $\nu=2,~\varOmega=1,~\alpha=7.16,~\omega=40,~\beta_l=1,~\beta_r=3,~\gamma=0.5$. In panels (a)-(d), the initial states are the same and taken as $|\varPsi \left( 0 \right) \rangle =\left( \frac{1}{\sqrt{2}},~0,~0,~\frac{1}{\sqrt{2}} \right) ^{\textrm{T}}$. (e)-(f) Parameters are the same as in (c)-(d), but with different
initial state $|\varPsi \left( 0 \right) \rangle =\left( \sqrt{0.4},~0,~0,~\sqrt{0.6} \right) ^{\textrm{T}}$.}
\label{figure6}
\end{figure}

To corroborate the above analytical results, the population imbalance and  magnetization (left column) and the atomic and spin currents (right column) are plotted versus time, by direct integration of the time-dependent Schr\"{o}dinger equation with Hamiltonian \eqref{con:1}. In figure \ref{figure6}, we consider two scenarios with $\gamma=0$ and $\gamma=0.5$, and take two sets of the parameter:  (a)-(b) $\nu=2,~\varOmega=1,~\alpha=22.52,~\omega=40,~\beta_l=1,~\beta_r=2,~\gamma=0$ and (c)-(d) $\nu=2,~\varOmega=1,~\alpha=7.16,~\omega=40,~\beta_l=1,~\beta_r=3,~\gamma=0.5$, to match the condition \eqref{con:35}. The initial states for both cases are taken as $|\varPsi \left( 0 \right) \rangle =\left( c_{l\uparrow},~c_{l\downarrow},~c_{r\uparrow},~c_{r\downarrow} \right) ^{\textrm{T}}=\left( \frac{1}{\sqrt{2}},~0,~0,~\frac{1}{\sqrt{2}} \right) ^{\textrm{T}}$. For both cases, we observe that after certain period of time, the population imbalances tend to  steady nonzero positive values which signal the spatial localization of the condensate in the amplifying well, whereas  magnetization exhibits periodic oscillation. As a consequence, the atomic currents tend to zero, while the spin currents are always nonzero, as illustrated in figures \ref{figure6} (b) and (d). In figure \ref{figure6} (e)-(f), we carry out the numerical studies of the populations and current behaviours with the same parameters as those in figure \ref{figure6} (c)-(d), but only with different initial state $|\varPsi \left( 0 \right) \rangle =\left( \sqrt{0.4},~0,~0,~\sqrt{0.6} \right) ^{\textrm{T}}$. Apparently, though the initial state is altered, the same physical effect (zero atomic current and nonzero spin current) can be attained, and the conclusion remains unaffected, except for the magnitudes of asymptotic population imbalance and oscillating amplitudes of magnetization (spin current). With the system parameters and initial states presented in figures \ref{figure6} (c) and (e), according to Eq.~\eqref{con:40}, we analytically calculate the asymptotic  values of  population imbalances, which read $0.63$ and $0.56$ as indicated by the
horizontal lines, showing good agreement with the numerical results.
Thus, we demonstrate, both analytically and  numerically, that the net spin current and steady state with spatial localization can be accessible in the dissipative Floquet spin-orbit-coupled system by adjusting the driving parameters.
\section{Conclusions}
We have analytically and numerically explored the populations and current behaviors of a single spin-orbit-coupled bosonic atom held in an open double well under periodic driving. Under the high-frequency driving, we have deduced the effective Hamiltonian  by using the time-averaging method, and obtained the analytical Floquet states and quasienergies of the considered system. We have explored the joint effects of SOC and periodic driving on  Josephson tunneling and current behaviors  in a non-Hermitian double-well system. Interesting, it is found that if the values of SOC strength are taken of half-integer numbers, no matter what values the other parameters are taken, the $\mathcal{P} \mathcal{T}$ phase transition can appear even  for arbitrarily small gain-loss coefficients.

In addition, we have revealed that the net spin current (zero atomic current and nonzero spin current) effect can not exist in the unbroken $\mathcal{PT}$-symmetric region, whereas it can survive in the broken $\mathcal{PT}$-symmetric region, if we drop the contribution of the growth of the norm  to the  current behaviors. In other words, in the broken $\mathcal{PT}$-symmetric region, the normalized atomic current can be zero while the normalized spin current can be nonzero at the same time. Nevertheless, the existence of net spin current (zero normalized atomic current and nonzero normalized  spin current) in the broken $\mathcal{PT}$-symmetric region is not a general feature for arbitrary values of SOC strength. When the dissipation is greater than the gain, we have found that the net spin current effect can be truly realized by tuning the parameters of the periodic driving field to match certain conditions. In such a non-$\mathcal{PT}$-symmetric system, the periodic driving enables
the system to approach a steady state accompanied with atomic localization phenomenon, for which the atomic current is zero while the spin current is nonzero.

\section{ACKNOWLEDGMENTS}
We thank Yunrong Luo for helpful discussions. This work was supported by the Natural Science Foundation of Zhejiang Province, China (Grant No. LY21A050002), the National
Natural Science Foundation of China (Grant No. 11975110), the Scientific and Technological Research Fund of
Jiangxi Provincial Education Department (Grants No. GJJ211026, GJJ190577 and No. GJJ190549), Zhejiang Sci-Tech University Scientific Research
Start-up Fund (Grant No. 20062318-Y), the Program for New Century Excellent Talents in University (Grant No. NCET-13-0836), the Scientific Research Foundation of Hunan Provincial Education Department (Grant No. 21B0063), and the Hunan Provincial Natural Science Foundation of China (Grant No. 2021JJ30435).

Jia Tang and Zhou Hu contributed equally to this work.

\section{References}

\end{document}